\documentclass{llncs}
\usepackage{llncsdoc}

\usepackage{multirow}
\usepackage{courier}
\usepackage{amsmath}
\usepackage{amsfonts}
\usepackage{subfig}
\usepackage{enumitem}
\usepackage{graphicx}
\usepackage{multicol}
\usepackage[labelfont=bf]{caption}
\usepackage{subfig}
\usepackage{float}
\usepackage{color}

\usepackage{booktabs} 
\usepackage{amsmath}
\usepackage{subfig}
\usepackage{amssymb}
\usepackage{graphicx}
\usepackage{enumitem}
\usepackage{textcomp }

\begin{document}
\title{Similar but Different: Exploiting Users' Congruity for Recommendation Systems}
\author{Ghazaleh Beigi \and Huan Liu}
\institute{\email{\{gbeigi, huan.liu\}@asu.edu} \\Arizona State University, Tempe, Arizona, USA}

\maketitle

\vspace{-10pt}
\begin{abstract}
	The pervasive use of social media provides massive data about individuals' online social activities and their social relations. The building block of most existing recommendation systems is the similarity between users with social relations, i.e., friends. While friendship ensures some homophily, the similarity of a user with her friends can vary as the number of friends increases. Research from sociology suggests that friends are more similar than strangers, but friends can have different interests. Exogenous information such as comments and ratings may help discern different degrees of agreement (i.e., congruity) among similar users. In this paper, we investigate if users' congruity can be incorporated into recommendation systems to improve it's performance. Experimental results demonstrate the effectiveness of embedding congruity related information into recommendation systems.
\end{abstract}

\vspace{-20pt}
\section{Introduction}
\vspace{-10pt}
Recommender systems play an important role in helping users find relevant and reliable information that is of potential interest~\cite{koren2010collaborative}. 
The increasing popularity of social media allows users to participate in online activities such as expressing opinions and emotions~\cite{beigi2016overview} (via commenting or rating), establishing social relations~\cite{gbeigi,beigi2016signed,beigi2016exploiting} and communities~\cite{alvari2016identifying}. 
Extracting these additional information (e.g., social relations) from social networks in favor of the task of recommendation, has attracted increasing attentions lately~\cite{ma2011recommender,jiang2012social}. In particular, homophily~\cite{mcpherson2001birds} which states that friends are more likely to share similar preferences with each other than strangers, is the backbone paradigms of recommendation systems that exploit social relations.

Despite the close-knit interests between friends, their friendship shall not always be treated as if they are completely alike. Naturally, as the number of friends of a user grows, it is inevitable that her friends' preferences diverge~\cite{Tang:2016:RSD:3015812.3015849,tang2009scalable}. For example, a user's friend circles are constituted of different people with various backgrounds and interests, ranging from her family to her schoolmates or co-workers. Research findings from sociology suggest that friends can make different decisions and many a time, these decisions can be very different from each other~\cite{cocking1998friendship}. Furthermore, although individuals have the tendency to become similar within a friendship, this should be considered as an effect of friendship, not a constitutive of it~\cite{cocking1998friendship}. Sociologists have also shown that the level of similarity among online users is much lower than that of actual friends in the real world~\cite{antheunis2012quality}. Thus, considering similarity at friendship level alone could be too coarse for recommendation tasks and might degrade the performance. 
\begin{figure}[ht]\vspace{-10pt}
\centering 
\subfloat[\bf{$u$'s local network.}]{\includegraphics[width=0.30\textwidth]{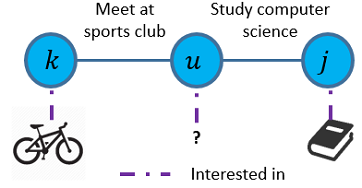}}\quad \quad \quad \quad
\subfloat[\bf{$u$'s opinion about others' interest.}]{\includegraphics[width=0.30\textwidth]{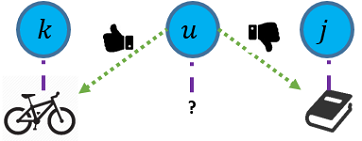}}
\caption{\textbf{The role of opinion agreement in inferring user $u$'s interests.}}\label{example}\vspace{-10pt}
\end{figure}

To give a palpable understanding of the above scenario, let us take a look at a toy example in Fig~\ref{example}. Assume user $u$ is connected to user $j$ since they both study computer science, and is connected to user $k$ because they usually meet at the same sports club (see Fig~\ref{example}(a)). Our goal is then to infer the interests of user $u$, given the information about interests of users $j$ and $k$. Social relations alone as representative of shared preferences would suggest that user $u$ is interested in both Machine Learning books and biking. However, from the opinions that user $u$ has expressed towards others' interests (see Fig~\ref{example}(b)), we can infer that she seems to be only interested in what user $k$ is interested in. Therefore, exogenous information such as the opinions of users regarding each other's interests can help inferring varying interests between them.

In this study, we use the term \textit{congruity} defined as a \textit{degree of agreement} between people~\cite{TCDP1999}, to refer to such a \textit{degree of match} among users' opinions. In other words, according to the sociology, congruity shall be treated as a perceptual concept that captures consensus between people~\cite{schein2010organizational,enz1988role}. Augmenting the recommendation systems with congruity might help precise inferring of the users' preferences. None of the existing recommendation systems until now have taken this into account.
The merit of exploiting the congruity obtained from users' interactions is that we can capture their preferences more accurately with potentials in improving the performance of recommendars, while it also poses new challenges. First, users' congruity information is not always readily available and extra effort is required to extract users' opinions towards each other's interests-- this is in contrast to the ideal scenario in the example shown in Fig~\ref{example}. Second, it's challenging to wisely incorporate congruity in recommendation systems.

The abundant information about users' behaviors and interactions is a rich source of users' congruity as most social media websites allow for free interaction and exposing viewpoints between users. This information has also potentials in distinguishing between congruity and social relations. In this paper, we seek to answer the following research questions: (1) What is the relationship between users' social relations (or friendship) and congruity? how different are social relations and congruity? (2) Why is it sensible to integrate congruity in recommendation systems? and (3) How can we mathematically obtain users' congruity from social data? We then propose a novel framework based on congruity for recommendation systems (CR). 
\vspace{-15pt}
\section{Data Analysis}
\vspace{-10pt}

We use two large online product-review websites, Epinions and Ciao, where users can establish friendship links toward each other from which we can construct the user-user social relations matrix $\mathbf{G}$. We denote by $\mathbf{G}_{ij}=1$, if $u_i$ and $u_j$ are friends, and $\mathbf{G}_{ij}=0$ otherwise. Different products are given ratings of 1 to 5 by users. From these ratings, we build our user-item rating matrix $\mathbf{R}$ where $\mathbf{R}_{ij}$ is the rating score that user $u_i$ has given to the item $v_j$. Users are also allowed to write reviews and can express their opinions toward each other by rating how helpful their reviews were from 1 to 5. Some key statistics are shown in Table~\ref{tab:data}. We perform some standard preprocessing by filtering out items and users with less than 3 ratings and users without social relations.
\begin{table}\vspace{-20pt}
	\centering
	\small
	\caption{\textbf{Statistics of the preprocessed data.}}\label{tab:data}
	\begin{tabular}{|l|l|l|}
		\hline
		\textbf{Name} & \textbf{Epinions} & \textbf{Ciao} \\ \hline\hline
		Users & 22,264 & 6,852\\ \hline
		Items & 35,040 & 16,202\\ \hline
		Ratings & 577,692 & 159,615\\ \hline
		Friendships & 292,345 & 111,672\\ \hline
		Pairs of Users with Congruity & 621,327 & 575,414\\ \hline
	\end{tabular}\vspace{-30pt}
\end{table}
\subsection{Congruity and Social Relations Difference}
\begin{figure}
	\centering 
	\includegraphics[width=0.32\textwidth]{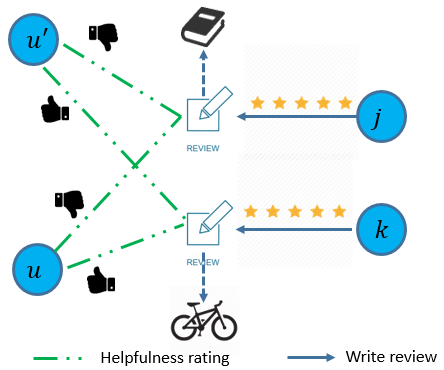}\\
	\caption{\textbf{An illustration of users' interaction in product review sites.}}\label{Example2}\vspace{-10pt}
\end{figure}

Recall that congruity is defined as a degree of agreement between people, which captures the socially defined levels of consensus between them~\cite{schein2010organizational,enz1988role}, and can be gleaned from users' interaction data. To illustrate this, let us glance at Fig~\ref{Example2} which demonstrates the typical users' interactions on websites such as Epinions and Ciao. Note, this is an extension to our toy example in the previous section, in that we have added another user $u'$ which is \textbf{not} connected to the existing users. In this example, users $u$ and $u'$ could rate the helpfulness of reviews written by users $k$ and $j$ on the bike and Machine Learning book. The high helpfulness rating that user $u$ has given to $k$'s review demonstrates the high level of opinion agreement and congruity between them, while the low helpfulness rating given to $j$'s review, by user $u$, implies lower congruity between them. Likewise, user $u'$ has similar congruity levels with users $k$ and $j$, however, we cannot infer the congruity level between $u$ and $u'$, given this information. 

Accordingly, we construct the user-user congruity matrix from users positive and negative interaction matrices, $\textbf{P}$ and $\textbf{N}$, which are obtained from helpfulness ratings as follows. First we consider high helpfulness ratings $\{4, 5\}$ as positive user interaction, low helpfulness ratings $\{1, 2\}$ as negative interaction and rating $\{3\}$ as neutral. Then, for each pair of users $\langle u_i,u_j\rangle$, we count the number of positive and negative interactions, $p_{ij}$ and $n_{ij}$, between $u_i$ and $u_j$. We calculate the positive interaction strength $\mathbf{P}_{ij}$ as a function of $p_{ij}$, i.e., $\mathbf{P}_{ij}=g(p_{ij})$ where $\mathbf{P}_{ij}\in[0,1]$. Therefore, we need the function $g(x)$ to have the following properties: (1) $g(0)=0$, (2) $\lim_{x\to\infty} g(x)=0$, and (3) be an increasing function of $x$. One choice could be $g(x)=	1-\frac{1}{\log(x+1)}$ for $x \neq 0$ and $g(x)=0$ otherwise.

Likewise, we construct the user-user negative interaction matrix $\mathbf{N}$. Ultimately, we create the user-user congruity matrix $\mathbf{C}$ by utilizing user-user positive and negative interaction matrices. Positive interactions imply more congruity between users while negative interactions imply the opposite.  Matrix $\mathbf{C}$ is then built from the linear combination of $\mathbf{P}$ and $\mathbf{N}$ as $\mathbf{C}=\mathbf{P}-\mathbf{N} $. 
 Note that there might be other ways to construct $\mathbf{C}$, $\mathbf{P}$ and $\mathbf{N}$, which we leave to future work.

Next, we further dig into our preprocessed data. 
 As we see from Table~\ref{tab:data_stat}, there are four possible types of pairs of users in our data: (1) users who are friends with each other and are congruent, (2) users who are friends with each other but are incongruent ($\langle i,j\rangle$ are incongruent if $\mathbf{C}_{ij}\leq 0$), (3) users who are strangers but are congruent, and (4) strangers who are also incongruent. These statistics suggest that, not all friends are always congruent. In particular, $24\%$ of friends in Ciao and $43\%$ of friends in Epinions are not congruent at all. Another interesting observation is that, $85\%$ and $73\%$ of pairs of congruent users in Ciao and Epinions are not friend with each other. Consequently, there might be some users with a degree of match in their preferences, who are not necessarily within their friend circles of each other. On the other hand, the number of congruent users in both datasets are much more than that of friends, which results in significantly different sets of users. These all, ultimately motivate us to exploit congruity for recommendation tasks rather than merely using social relation.
\begin{table}[h]
	\vspace{-22pt}
	\centering
	\small
	\caption{\textbf{Number of pairs of users with different properties.
			}}\label{tab:data_stat}
	\subfloat[\bf{Ciao}]{
		\begin{tabular}{|l|l|l|}
			\hline
			& \textbf{Congruent} & \textbf{Incongruent} \\ \hline
			\textbf{Friends} & 84,063 & 27,609 \\ \hline
			\textbf{Strangers} & 491,351 & $\sim$ 46 M \\ \hline
		\end{tabular}}
		\quad
		\subfloat[\bf{Epinions}]{
			\begin{tabular}{|l|l|l|}
				\hline
				& \textbf{Congruent} & \textbf{Incongruent} \\ \hline
				\textbf{Friends} & 163,985 & 128,360 \\ \hline
				\textbf{Strangers} & 457,342 & $\sim$ 494 M\\ \hline
			\end{tabular}}\vspace{-20pt}
		\end{table}
		\vspace{-22pt}
\subsection{Analysis of Users' Congruity}
Before leveraging users' congruity for recommendation tasks, we would like to conduct a sanity check to see if this concept is applicable to social media data. We first study if social relations between users correspond to their congruency or in other words, if all friends are congruent with each other or not. Then, we investigate the correlation between users' congruity and preferences.  Specifically, we verify two questions: (1) Are all friends congruent?, and (2) Does congruity among users imply a higher chance of sharing similar preferences between them? 



To answer the first question, for each user $u_i$, we consider all of her friends. Then, we compute the minimum $c^{i}_{min}$ and maximum $c^{i}_{max}$ values of congruity between user $u_i$ and her friends. Two vectors $\mathbf{c}^{min}$ and $\mathbf{c}^{max}$ are obtained by computing $c^{min}$s and $c^{max}$s for all users. We conduct a two-sample t-test on $\{\mathbf{c}_{min},\mathbf{c}_{max}\}$ where the null hypothesis $H_0$ is that friends are all congruent, i.e. there is no significant difference between minimum and maximum value of users' congruity. The $H_1$ is also that friends are not all congruent:\vspace{-5pt}
\begin{equation}
H_0: \mathbf{c}_{min} = \mathbf{c}_{max},~~~~~H_1: \mathbf{c}_{min} \neq\mathbf{c}_{max}.\vspace{-5pt}
\end{equation}

The null hypothesis is rejected at significance level $\alpha=0.01$ with p-value shown in Table~\ref{pvalue}. This suggests a negative answer to the first question.

A similar procedure we did for the first question can be followed to answer the second question. Consider the pair of user $\langle u_i,u_j\rangle$ with positive value of congruity ($\mathbf{C}_{ij}>0$). We randomly select user $u_k$ who is incongruent with $u_i$. Users similarities $cp^{ij}$ and $cr^{ik}$ have been calculated for $\langle u_i,u_j\rangle$ and $\langle u_i,u_k\rangle$, respectively. Finally, two vectors $\mathbf{c}_p$ and $\mathbf{c}_r$ are obtained where $\mathbf{c}_p$ is the set of all $cp$'s for pairs of users with congruity; while $\mathbf{c}_r$ is the set of $cr$'s for pairs of users without congruity. We use cosine similarity over the item-rating entries to find the similarities between users. We conduct a two-sample t-test on $\{\mathbf{c}_p,\mathbf{c}_r\}$ where the null hypothesis $H_0$ is that users without congruity are more likely to share similar preferences:\vspace{-5pt}
\begin{equation}
H_0: \mathbf{c}_p\leq \mathbf{c}_r,~~~~~H_1: \mathbf{c}_p> \mathbf{c}_r.\vspace{-5pt}
\end{equation}

The null hypothesis is rejected at significance level $\alpha=0.01$. Thus, users with congruity are more likely to share preferences than those without.

The corresponding p-values for the above t-tests are summarized in Table~\ref{pvalue} for both datasets. The results from these analyses: 1) demonstrate that although friends share similar interests, but friendship relations are not good measure of congruency as friends are not always congruent with each other and thus considering similarity at friendship level alone degrade the performance of recommendation tasks, and 2) confirm the importance of deploying a measure of users' congruity in computing users' similarity other than social relations. 
\begin{table}
	\vspace{-20pt}
	\centering
	\small
	\caption{\textbf{p-values of t-test results corresponding to analysis tests.}}\label{pvalue}
	\begin{tabular}{|l|l|l|l|l|}\hline
		 &Ciao: $\{\mathbf{c}_{min},\mathbf{c}_{max}\}$ & Ciao: $\{\textbf{c}_p,\textbf{c}_r\}$ & Epinions: $\{\mathbf{c}_{min},\mathbf{c}_{max}\}$ & Epinions: $\{\textbf{c}_p,\textbf{c}_r\}$\\ \hline
		 p-value& 3.09e--6 & 1.72e--5 & 6.17e--5 & 4.81e--4 \\ \hline
	\end{tabular}\vspace{-20pt}
\end{table}
\vspace{-17pt}
\section{Congruity-Based Recommendation}
\vspace{-5pt}
We begin this section by introducing matrix factorization based collaborative filtering technique which we chose as basis of CR. Matrix factorization based techniques have been widely used for building recommender systems~\cite{koren2009matrix,ma2011recommender} and the basic assumption is that a small number of factors influence user rating behavior and maps both user and item to a joint latent factor space with dimensionality $d$. Assume that $\mathbf{U}_i \in \mathbb{R}^{1 \times d}$ and $\mathbf{V}_j\in\mathbb{R}^{1 \times d}$ are the user preference vector for $u_i$ and item characteristic vector for $v_j$, respectively. The rating score given by $u_i$ to $v_j$ is modeled as $\mathbf{R}_{ij} = \mathbf{U}_i\mathbf{V}_j^\top$. Matrix factorization seeks to find $\mathbf{U}=[\mathbf{U}_1,...,\mathbf{U}_n]$ and $\mathbf{V}=[\mathbf{V}_1,...,\mathbf{V}_m]$ by solving the following problem:
\begin{equation}\label{SimpleMF}
\min_{\mathbf{U,V}}~~\sum_{i=1}^{n}\sum_{j=1}^{m} \mathbf{I}_{ij}(\mathbf{R}_{ij}-\mathbf{U}_i\mathbf{V}_j^\top)^2 +\lambda(\|\mathbf{U}\|^2_\mathbf{F}+\|\mathbf{V}\|^2_\mathbf{F})
\end{equation}
\noindent where $\lambda(\|\mathbf{U}\|^2_\mathbf{F}+\|\mathbf{V}\|^2_\mathbf{F})$ is added to avoid over-fitting and $\mathbf{I}_{ij}$ controls the contribution from $\mathbf{R}_{ij}$. A typical choice of ${\mathbf I}$ is ${\bf I}_{ij}=1$ if ${\bf R}_{ij}\neq 0$ and ${\bf I}_{ij}=0$, otherwise.
The observations in the previous section demonstrated that although friends are similar, they are not all congruent and we need to integrate congruity to measure the similarity of shared preferences between users. Moreover, users with congruity are more likely to share similar preferences compared to those without either of them. These findings provide the groundwork for us to model users' congruity to measure the users' preferences closeness.

Let $\mathbf{L}\in \mathbb{R}^{n\times n}$ be the preference closeness matrix where $\mathbf{L}_{ik}$ denotes the preference closeness strength between $u_i$ and $u_k$. The motivation behind preference closeness strength is that users are more/less likely to share similar preferences when they establish higher/lower level of congruity with each other. Following this idea, $\mathbf{L}_{ik}$ could be calculated as the congruity $\mathbf{C}_{ik}$ between them. The closeness of $u_i$ and $u_k$ user preference vectors is then controlled by their preference closeness strength,\vspace{-5pt}
\begin{equation}
\label{belief}
\min \sum_{i=1}^{n} \sum_{k\in \mathcal{T}_i}^{} \mathbf{L}_{ik}\|\mathbf{U}_i - \mathbf{U}_k\|^2_2
\end{equation}
\noindent where $\mathcal{T}_i=\{u_k|\mathbf{C(i,k)}\neq 0\}$. In Eq.~\ref{belief}, a larger value of $\mathbf{L}_{ik}$ indicates the strong association between $u_i$ and $u_k$; hence $u_i$'s preference vector $\mathbf{U}_i$ is more likely to be close to $u_k$'s preference vector $\mathbf{U}_k$-- this makes the distance between $\mathbf{U}_i$ and $\mathbf{U}_k$, smaller. While a smaller value of $\mathbf{L}_{ik}$ indicates weak association between $\mathbf{U}_i$ and $\mathbf{U}_k$; therefore their distance is larger. Having introduced our solutions to model users' congruity, our framework, congruity based recommendation system (CR), is to minimize the following problem,
\begin{align}
\label{formula:Opt1}
\mathcal{J} =\min_{\mathbf{U,V}}~~\sum_{i=1}^{n}\sum_{j=1}^{m}\mathbf{I}_{ij}(\mathbf{R}_{ij}-\mathbf{U}_i\mathbf{V}_j^\top)^2
+\gamma \sum_{i=1}^{n} \sum_{k\in \mathcal{T}_i}^{} \mathbf{L}_{ik}\|\mathbf{U}_i - \mathbf{U}_k\|^2_2 +\lambda(\|\mathbf{U}\|^2_\mathbf{F}+\|\mathbf{V}\|^2_\mathbf{F})
\end{align}
\noindent where $\gamma$ is used to control contributions of the users' preferences closeness strength. We use gradient descent method to solve Eq.~\ref{formula:Opt1}, which has been proven to gain an efficient solution in practice. The partial derivations of $\mathcal{J}$ with respect to $\mathbf{U}_i$ and $\mathbf{V}_j$ are as follows,
\begin{align}\label{formula:devUi}
	\frac{1}{2}\frac{\partial {\mathcal{J}}}{\partial \mathbf{U}_{i}} =  -\sum_{j}\mathbf{I}_{ij}(\mathbf{R}_{ij}-\mathbf{U}_{i}\mathbf{V}_j^\top)\mathbf{V}_j + \lambda \mathbf{U}_{i}+\gamma\sum_{k \in \mathcal{T}_i}\mathbf{L}_{ik}(\mathbf{U}_{i}-\mathbf{U}_{k})
\end{align}\vspace{-15pt}
\begin{align}\label{formula:devVj}
&\frac{1}{2}\frac{\partial {\mathcal{J}}}{\partial \mathbf{V}_{j}} = -\sum_{i}\mathbf{I}_{ij}(\mathbf{R}_{ij}-\mathbf{U}_{i}\mathbf{V}_j^\top)\mathbf{U}_i + \lambda \mathbf{V}_{j}
\end{align}\vspace{-5pt}

We use Eq.~\ref{formula:devUi} and Eq.~\ref{formula:devVj} to update $\mathbf{U}$ and $\mathbf{V}$ until convergence. After learning the user preference matrix $\mathbf{U}$ and the item characteristic matrix $\mathbf{V}$, an unknown score $\hat{\mathbf{R}}_{i'j'}$ from the user $u_{i'}$ to the item $v_{j'}$ will be predicted as $\hat{\mathbf{R}}_{i'j'} = u_{i'}^\top v_{j'}$.
\vspace{-10pt}
\section{Experiments}
\vspace{-10pt}
In this section, we conduct experiments to answer the following two questions: (1) Does leveraging users' congruity help recommendation?, and (2) How does integration of users' congruity with social relations improve recommendation performance? and which one of the congruity and social relations contribute most to the performance improvement?

We use Root Mean Square Error (RMSE) and Mean Absolute Error (MAE) to evaluate the performance and smaller values indicate the better performance. Note that \textit{small improvement in RMSE or MAE terms could result in a significant impact on the quality of top few recommendations}~\cite{koren2008factorization}. In this work, we randomly select $x\%$ of the ratings as training and treat the remaining $(100-x)\%$ as test ratings to be predicted. We vary $x$ as $\{40, 50, 70, 90\}$.
\vspace{-15pt}
\subsection{Performance Comparison}
\vspace{-5pt}
To answer the first question, we compare the proposed framework CR with the following recommender systems:
\begin{itemize}[leftmargin=*]
\item \textbf{MF}: It performs basic matrix factorization on the user-item rating matrix to predict the new ratings by only utilizing the rating information~\cite{salakhutdinov2008probabilistic} 
\item \textbf{SMF}: Similarity based matrix factorization method is a variation of our method which uses user-user similarity matrix $\mathbf{S}$ for calculating $\mathbf{L}$. We use cosine similarity over the item-rating entries to find the users similarities.
\item \textbf{SoReg}: It performs matrix factorization while exploiting social regularization defined based on both user-item matrix and positive social relations~\cite{ma2011recommender}.
\item \textbf{DualRec}: It integrates both review and rater roles of each user and uses item and review helpfulness ratings to learn reviewer and rater roles, respectively.
\end{itemize}

It's notable to say that other social relation based recommenders such as~\cite{ma2009learning,ma2009ste} have comparable results with~\cite{ma2011recommender}. Note that \textit{CR} incorporates users' congruity while all three baselines \textit{SoReg}, \textit{DualRec} and \textit{SMF} methods use social relations, helpfulness ratings and user-user rating similarity, respectively. This results in a substantially different method in terms of both key ideas and techniques. For all baselines with parameters, we use cross-validation to determine their values. For the proposed framework we set the parameters for Epinions and Ciao $\{\lambda =0.01, \gamma = 100, d = 15\}$, and $\{\lambda =0.01, \gamma = 10, d = 20\}$, respectively. Since the test set is selected randomly, the final results are reported by taking the average of 20 runs for each method. We also conduct a t-test on all comparisons, and the results are significant. The comparisons on Epinions are shown in Table~\ref{Res_Ci}. We also conduct experiments on Ciao and observe very similar trends. Due to lack of space, we leave the results out. We have the following observations,

\begin{itemize}[leftmargin=*]
\item All methods outperform \textit{MF}, suggesting the importance of leveraging exogenous information (e.g. users' congruity, social relations and rater roles) for improving the performance of recommendation systems.
\item \textit{SMF} fails to demonstrate comparable results compared to all other methods. This indicates that users' rating similarity cannot capture their shared preferences as good as social relations and congruity.
\item The proposed framework \textit{CR} always obtains the best performance. The reason is that using social relations or rater roles of users does not capture the closeness of users' preferences. This confirms the effectiveness of users' congruity in learning their preferences and improving the performance of recommenders. 
\end{itemize}
\begin{table*}[t]\vspace{-30pt}
	\caption{\textbf{Performance comparison of different methods.}}\label{Res_Ci}
	\centering 
	\small
	\begin{tabular}{|c | c | c | c | c | c | c |}
		\hline
		Training & Metrics & \textbf{MF} & \textbf{SMF} & \textbf{SoReg} & \textbf{DualRec} & \textbf{CR}\\ \hline
		\multirow{2}{20pt}{90\%} & MAE & 0.9768$\pm0.0027$ & 0.9578$\pm0.0028$ & 0.9352$\pm0.0030$ & 0.9231$\pm0.0026$ & $\mathbf{0.9136\pm0.0029}$\\
		& RMSE & 1.1687$\pm0.0029$ & 1.1476$\pm0.0027$ & 1.1294$\pm0.0030$ & 1.1167$\pm0.0028$ & $\mathbf{1.1041\pm0.0032}$ \\\hline
		\multirow{2}{20pt}{70\%} & MAE & 0.9848$\pm0.0029$ & 0.9611$\pm0.0031$ & 0.9417$\pm0.0027$ & 0.9387$\pm0.0030$ & $\mathbf{0.9252\pm0.0034}$ \\
		& RMSE & 1.1776$\pm0.0030$ & 1.1597$\pm0.0029$ & 1.1356$\pm0.0031$ & 1.1253$\pm0.0028$ & $\mathbf{1.1172\pm0.0030}$ \\\hline
		\multirow{2}{20pt}{50\%} & MAE & 0.9921$\pm0.0026$ & 0.9702$\pm0.0027$ & 0.9539$\pm0.0029$ & 0.9471$\pm0.0029$ & $\mathbf{0.9335\pm0.0031}$ \\
		& RMSE & 1.1894$\pm0.0031$ & 1.1655$\pm0.0030$ & 1.1478$\pm0.0031$ & 1.1416$\pm0.0030$ & $\mathbf{1.1339\pm0.0034}$ \\\hline
		\multirow{2}{20pt}{40\%} & MAE & 0.9969$\pm0.0029$ & 0.9783$\pm0.0030$ & 0.9582$\pm0.0029$ & 0.9506 $\pm0.0029$ &$\mathbf{0.9400\pm0.0035}$ \\
		& RMSE & 1.1932$\pm0.0025$ & 1.1761$\pm0.0028$ & 1.1531$\pm0.0029$ & 1.1446$\pm0.0029$ & $\mathbf{1.1378\pm0.0031}$ \\\hline 
	\end{tabular} \vspace{-10pt}
\end{table*}
\vspace{-12pt}
\subsection{Integrating Congruity with Social Relations}
\begin{figure*}[t]
	\centering 
	\subfloat[\bf{RMSE}]{\includegraphics[width=0.37\textwidth]{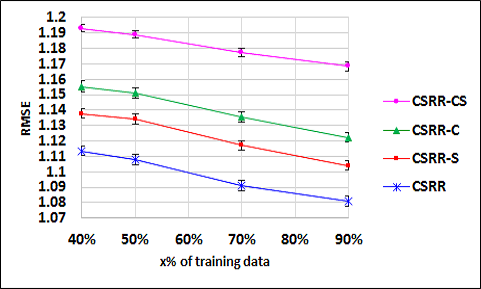}}\quad
	\subfloat[\bf{MAE}]{\includegraphics[width=0.37\textwidth]{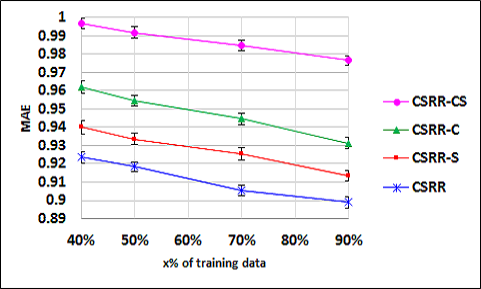}}
	\caption{\textbf{Effect of social relations and congruity in Epinions.}}\label{Res_impact1}\vspace{-15pt}
\end{figure*}

\vspace{-8pt}
Here, we investigate the impact of integrating congruity along with social relations and then study the effect of each to answer the second question. To achieve this goal, we define \textbf{CSRR} as a variation of our proposed method in which the preference closeness matrix  $\mathbf{L}\in \mathbb{R}^{n\times n}$ is updated as follows:\vspace{-5pt}
\begin{equation}\label{closen}
\mathbf{L}_{ik}=
\delta \mathbf{G}_{ik} + (1-\delta) \mathbf{C}_{ik}.
\vspace{-5pt}
\end{equation}
As discussed earlier, $\mathbf{G}_{ik}\in [0,1]$ and $\mathbf{C}_{ik}\in [-1,1]$, therefore, to reduce the further complexities in our model, we replace $\mathbf{C}_{ik}$ by $\frac{\mathbf{C}_{ij}+1}{2}\in [0,1]$. This further makes $\mathbf{L}_{ik}$ to be in $[0,1]$. 
Also $\delta$ controls the contributions of $\mathbf{G}_{ik}$ and $\mathbf{C}_{ik}$. Here, we set $\delta = 0.3$. We now define the following variants of CSRR as follows:
\begin{itemize}[leftmargin=*]
	\item \textit{CSRR--S}: Eliminates the effect of social relations by setting $\delta =0$ in Eq.~\ref{closen}. This variation is equal to the \textbf{CR} method described earlier;
	\item \textit{CSRR--C}: Eliminates the effect of congruity by setting $\delta = 1$ in Eq.~\ref{closen};
	\item \textit{CSRR--CS}: Eliminates the effects of both social relations and congruity by setting $\gamma =0$ in Eq.~\ref{formula:Opt1}. This variation is equal to the \textbf{MF} method.
\end{itemize}
The results are shown in Fig~\ref{Res_impact1} for Epinions. The results for Ciao have very similar trends but they are omitted due to space limit. We observe the following:
\vspace{-4pt}
\begin{itemize}[leftmargin=*]
	\item When we remove the effect of congruity, the performance of CSRR--C degrades compared to CSRR. We have the similar observations for the elimination of social relations. Those results support the importance of integrating congruity as well as social relations information in a recommender system.
	\item We note that the performance degrades more by eliminating congruity information, CSRR--C compared to eliminating social relations, CSRR--S. This is because the congruity information is much denser than social relations.
	\item Removing the effects of social relations and congruity, the performance of CSRR--CS reduces compared to CSRR--C and CSRR--S. This suggests that incorporating users' congruity along with social relations are important and have a complementary role to each other.
\end{itemize}
\vspace{-5pt}

To recap, users' congruity and social relations are two different sets of information. Exploiting congruity information has potentials in more accurate measuring of users' opinions degree of match.
\vspace{-8pt}
\section{Related Work}
\vspace{-10pt}
Collaborative filtering methods are categorized into the neighborhood-based and model-based models. 
The low-rank matrix factorization methods are one example of model-based methods which estimate the user-item rating matrix using low-rank approximations method to predict ratings~\cite{salakhutdinov2008probabilistic,koren2009matrix}. 
The increasing popularity of social media encourages individuals to participate in various activities which could provide multiple sources of information to improve recommender systems. Some algorithms incorporate user profile~\cite{park2009pairwise,zhou2011functional,wang2015leveraging}. For example, the work of~\cite{park2009pairwise} constructs tensor profiles of user-item pairs while the method in~\cite{zhou2011functional} makes a profile for users using the initial interview process to solve the cold-start problem in the recommendation. Another method~\cite{wang2015toward} considers both rater and reviewer roles of each user to improve recommendation. It uses item ratings and helpfulness ratings to obtain the reviewer and rater roles, respectively. 

Social relations also provide an independent source of information which brings new opportunities for recommendation~\cite{beigi2016exploiting,beigi2016signed,gbeigi,Tang:2016:RSD:3015812.3015849,ma2011recommender,ma2009ste,jamali2009trustwalker}. The work of~\cite{ma2011recommender} incorporates user social relations which force a user's preferences to be close to her friends' and is controlled by their similarity which is measured based on item-ratings. The work of~\cite{jamali2009trustwalker} proposes a random walk model based on users' friendship relations and item-based recommendation. The length of the random walk which is based on both item ratings and social relations. Another work~\cite{ma2009ste} proposes a probabilistic framework which assumes that individuals preferences could be influenced by their friends' tastes. It fuses both users' preferences and their friends' tastes together to predict the users' favors on items. 
The importance of exploiting heterogeneity of social relations~\cite{tang2009scalable} and weak dependency connections for recommendation systems has been shown in~\cite{Tang:2016:RSD:3015812.3015849}. To capture the heterogeneity of social relations and weak dependency connections, it adopts social dimensions by finding the overlapped communities in the social network. 

The difference between CR and the above models is that we investigate the role of users' congruity as social relations alone do not demonstrate the degree of opinion match between users. 
 Moreover, congruity is determined independently from social relationship information and is obtained from users' interactions to further capture different degrees of match between their opinions. 
 \vspace{-13pt}
\section{Conclusion and Future Work}
\vspace{-10pt}
In this paper, the concept of congruity, a degree of agreement and appropriateness between people, borrowed from sociology is tailored to discern different degrees of match between their opinions and enhance the performance of recommendation systems. To overcome the challenge that users' congruity is not readily available, we leverage the available users' interaction data and capture the congruity between users from data. 
We propose the framework CR, which predicts unknown user-item ratings by incorporating congruity information. We conduct experiments on real-world data, and the results confirm the efficiency of congruity for inferring users' opinions degree of match. 
In future, we would like to incorporate temporal information to study the dynamics of users' congruity in recommendation systems. Also, the findings of this work  may be helpful for other tasks such as friend recommendation in social networks. 
\vspace{-15pt}
\section{Acknowledgments}
\vspace{-8pt}
This material is based upon the work supported by, or in part by, Office of Naval Research (ONR) under grant number N00014-17-1-2605.

\end{document}